\documentstyle[12pt,eqsecnum,aps,epsf]{revtex}
\newcommand{\etal}{\mbox{\it et al.}}
\tighten

\begin{document}

\preprint{~}

\title{Observation of the Askaryan Effect: \\
Coherent Microwave Cherenkov Emission\\
 from Charge Asymmetry in High Energy Particle Cascades}

\author{
David~Saltzberg$^{1}$, 
Peter~Gorham$^{2}$,
Dieter Walz$^{3}$,
Clive Field$^{3}$,
Richard Iverson$^{3}$,
Allen Odian$^{3}$,
George Resch$^{2}$,
Paul~Schoessow$^{4}$,
and Dawn Williams$^{1}$}
\address{$^{1}$Department of Physics and Astronomy,
University of California, Los Angeles, CA 90095}
\address{$^{2}$Jet Propulsion Laboratory, Calif. Institute of Technology,
Pasadena, CA, 91109}
\address{$^{3}$Stanford Linear Accelerator Center, Stanford University, Stanford, CA 94309}
\address{$^{4}$Argonne National Laboratory, Argonne, IL}

\date{\today}

\maketitle

\begin{abstract}

We present the first direct experimental evidence for the charge
excess in high energy particle showers
predicted nearly 40 years ago by Askaryan. 
We directed bremsstrahlung photons from  picosecond
pulses of 28.5 GeV electrons at the SLAC Final Focus Test Beam
facility into a 3.5 ton silica sand target, producing
electromagnetic showers several meters long.
A series of antennas spanning 0.3 to 6 GHz 
were used to detect strong, sub-nanosecond
radio frequency pulses produced whenever a shower was present.
The measured electric field strengths are consistent
with a completely coherent radiation process. 
The pulses show 100\% linear polarization, consistent
with the expectations of Cherenkov radiation.
The field strength versus depth closely follows 
the expected particle number density profile of the cascade,
consistent with emission from excess charge distributed along the
shower. These measurements therefore provide strong support
for experiments designed to detect high energy cosmic rays and
neutrinos via coherent radio emission from their cascades.

\end{abstract}


\section{Introduction}
During the development of a high-energy electromagnetic cascade 
in normal matter, photon and electron scattering processes 
pull electrons from the surrounding material into 
the shower.  In addition, positrons in the shower
annihilate in flight.  The
combination of these processes should lead 
to a net 20-30\% negative charge excess for the comoving
compact body of particles that carry most of the shower energy.
G. A. Askaryan\cite{Ask62} first described this effect, and
noted that it should lead to strong coherent radio and microwave
Cherenkov emission
for showers that propagate within a dielectric. The range of
wavelengths over which coherence obtains depends on the 
form factor of the shower bunch---wavelengths shorter than
the bunch length suffer from
destructive interference and coherence is lost. 
However, in the fully coherent regime the radiated
energy scales quadratically with the net charge of the
particle bunch, and at ultra high energies the resulting coherent
radio emission may carry off a significant fraction of the total
energy in the cascade.

The plausibility of
Askaryan's arguments combined with more recent modeling and
analysis\cite{Mark86,ZHS92,Alv96,Alv97,Alv98} 
has led to a number of experimental searches for 
high energy neutrinos by exploiting the effect at energies 
from $\sim 10^{16}$ eV in Antarctic ice\cite{fri96,Bes99}
up to $10^{20}$ eV
or more in the lunar regolith, using large
ground-based radio telescopes\cite{Zhe88,Dag89,Han96,Gor99}.
Radio frequency pulses have been observed for many years from
extensive air showers\cite{Jel66,Feg68}. However, it has been
shown\cite{KL66,All71} that the most likely source of this emission
is a form of Lorentz-boosted dipole radiation from geomagnetic
charge separation in the air shower, rather than the
Askaryan effect. Thus neither the charge asymmetry
nor the resulting coherent Cherenkov radiation has ever been
observed.

In a previous paper\cite{P1} 
we have described initial efforts to measure the coherent
radio-frequency (RF) emission from electron bunches interacting in a solid 
dielectric target consisting of 360~kg of silica sand.  That study, done 
with relatively low-energy electrons (15 MeV), 
demonstrated the presence of coherent 
radiation in the form of extremely short and intense microwave 
pulses detectable over a wide frequency range. 
These results, while useful
for understanding the coherent RF emission processes from 
relativistic charged particles,
could not directly test the development of a shower charge excess,
as Askaryan predicted. Also, because the particles were charged,
passage of the beam through any interface
induced strong RF transition radiation (TR), which obscured the
presence of Cherenkov radiation (CR).

We report here on measurements made at the Stanford Linear Accelerator
Center (SLAC) in which the use of high-energy photons, rather than 
low-energy electrons, has enabled us to clearly observe
microwave Cherenkov radiation from the Askaryan effect.
The electromagnetic showers thus
produced in our target yielded strong, coherent, sub-ns RF pulses,
consistent in every way observed with the predictions. As we will show here,
our results conclusively support the reported sensitivity of the
experiments noted above.

\section{Experimental setup}

A silica sand
target  and associated antennas were placed in a
gamma-ray beamline in the Final Focus Test Beam facility at SLAC
in August 2000.  The apparatus was placed 30~m downstream of 
bremsstrahlung radiators that produced a high-energy photon
beam from 28.5~GeV electrons.
Just downstream of the bremsstrahlung targets, the electron
beam was bent down, passed 0.5~m 
below our sand target within a vacuum pipe,
and was dumped 15 m beyond the end of
our target. 
Typical beam currents during the experiment were
$(0.2-1.0) \times 10^{10}$ electrons per bunch. 
The two radiators could
be used either separately or in tandem, thereby providing
0\%, 1\%, 2.7\% or 3.7\% of a radiation length.
Thus the effective shower energy induced by
the photons
could be varied by a factor of $\sim 18$, from 
$(0.06-1.1) \times 10^{19}~$~eV per bunch.
The size of the photon bunch at the entrance of the
silica sand target was less than several~mm in all dimensions.

\begin{figure}
\begin{center}
\leavevmode
\epsfxsize=5in
\epsfbox{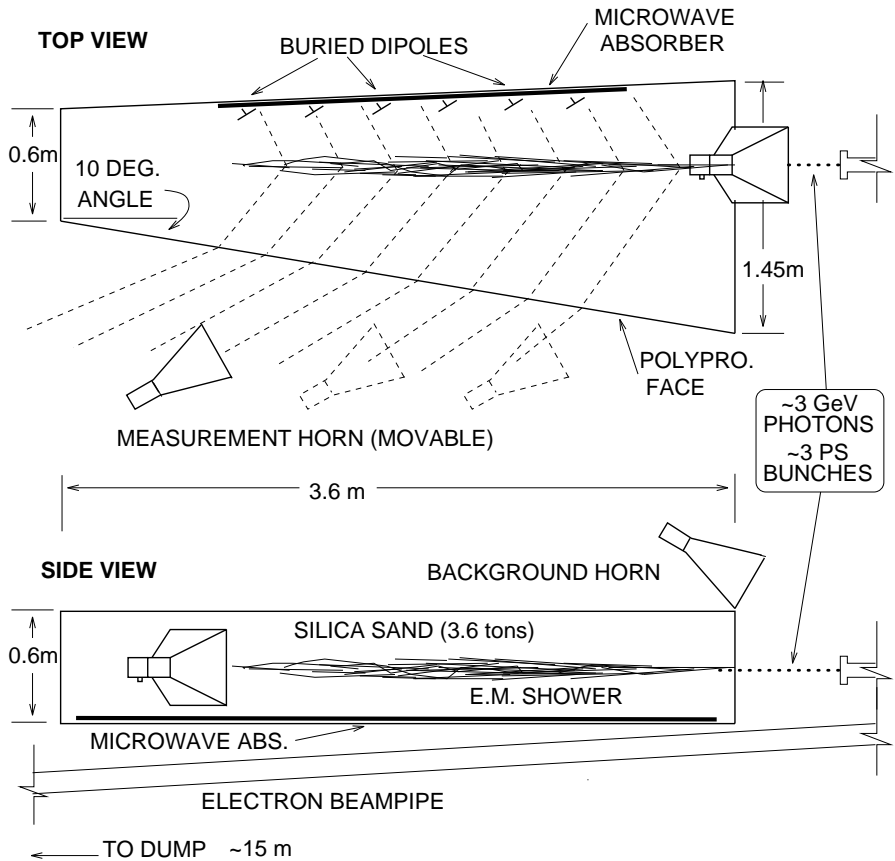}
\caption{Sectional views of the target geometry.}
\label{tgeom}
\end{center}
\end{figure}

The target was a large container built largely from
non-conductive materials
such as wood and plastic which we filled with 3200 kg of dry
silica sand.
As shown in Fig. 1,
the sand target was
rectangular in cross section perpendicular to the beam axis,
but the vertical faces on both sides were angled slightly to
facilitate transmission of radiation arriving at the Cherenkov angle
(about $51^{\circ}$ in silica sand at microwave frequencies). 
We avoided making sides
parallel to the beam since the CR would then suffer
total internal reflection at the interface. 
Both internal (buried) half-wave dipoles and external antennas were used
for the pulse measurements.  The external antennas were 
pyramidal ``standard gain'' microwave horns  (1.7--2.6~GHz or
4.4--5.6~GHz).
A low-microwave-loss
plastic (0.5 inch thick polypropylene) was used for the face
viewed by the external antennas.

Details of the trigger system, data acquisition, and polarization and 
power measurements were similar to those used in our previous
experiment\cite{P1}.  
Briefly, time-domain
sampling of the antenna voltages using high-bandwidth oscilloscopes
allowed us to make direct measurements
of the electric field of the pulses as a function of time, with
time resolutions of 100~ps (0.2--3 GHz) and 10~ps (4--6 GHz).
Because of the strength
of the signals no amplification was required.
Microwave
absorber material was placed wherever possible to minimize reflections,
and the geometry was chosen so that, where reflections could not
be eliminated, they would arrive well after the
expected main pulse envelope.  The incident electron 
beam current was measured using
a beam current transformer.

Because the accelerator itself uses S-band (2.9~GHz)
RF for the beam generation, we took
particular care to measure the background levels of RF with the
electron beam on, but no photon 
radiators in place.
We found a weak background RF pulse
of about 20~ns duration,
in coincidence with the beam,
at the level of a few~mV r.m.s.  
However, this background proved to be
completely negligible compared to the signal pulses
detected with the photon beam directed into the target, typically
10-100~V r.m.s.  In
addition, during most of the data runs, we used a second S-band horn
to monitor any incoming radiation just upstream of the target
(see fig. 1). These and numerous other tests eliminated 
the possibility that stray linac
RF or other RF associated with the electron beam contributed to 
our measurements.

We checked the static magnetic field strength in the
vicinity of our sand target and found it to be comparable to
ambient geomagnetic levels, of order 0.5~Gauss. This is important
since stray static fields present in the accelerator vault could have
induced significant charge separation in showers within our target and lead
to other possible radiation mechanisms. In a field of strength
0.5~Gauss, the electron gyroradius for the bulk of the
shower is $\geq 1$~km, implying charge separations of 
$\leq 1$~cm over the length of a shower, well below the $\sim 4$~cm
Moli\`{e}re radius. 

\section{Results}

As noted above, we observed strong RF pulses correlated in time to
the presence of a shower at all of our observed radio frequencies.
Having first eliminated the possibility that these were
simply due to a background process, the problem of establishing and
characterizing
their origin remains.
We report here on several measurements,
each of which provides independent information in determining
the source of the observed RF radiation. These are:
the pulse properties in the time domain,
the correlation of
measured pulse field strengths to the expected shower profile,
the polarization properties of the pulses,
and the spectral and beam current dependence of the
pulse field strengths.

\subsection{Time \& shower profiles}
The inset
to fig. 2 shows a typical pulse profile measured with
one of the S-band horns aimed near shower maximum. Given that
the bandwidth of the horn is 900~MHz, the $\sim 1$~ns 
pulse width indicates a bandwidth-limited signal. This behavior is
seen at all frequencies observed, with the best upper limit
to the intrinsic pulse width, 
based on the 4.4--5.6~GHz data, being less than 500~ps.

Fig. 2 also shows a set of measured peak field strengths for pulses
taken at different points along the shower. In this case the horn
was translated parallel to the shower axis, maintaining the
same angle (matched to the refracted Cherenkov angle) at each point.
The values are plotted at a shower position corresponding to
the center of the antenna beam pattern, refracted onto the shower.
The half-power beamwidth of the horn is about $\pm 10^{\circ}$;
thus the antenna would respond to any isotropic shower
radiation over a range of $\pm 20-25$~cm around the plotted positions.
For CR, the radiation pattern is expected to have a total beamwidth
of several degrees or less\cite{ZHS92,Tak94}, 
much smaller than the antenna beam pattern. 

The plotted curve shows the expected profile of the total
number of particles in the shower, based on the 
Kamata-Nishimura-Greisen (KNG)\cite{KNG} approximation. 
Here the field strengths have been scaled in the
plot to provide an approximate overlay to the relative shower profile.
Clearly the pulse strengths are highly correlated to the particle
number profile. Since the excess charge is also expected to closely follow
the shower profile, this result is consistent with Askaryan's hypothesis.

\begin{figure}
\begin{center}
\leavevmode
\epsfxsize=5in
\epsfbox{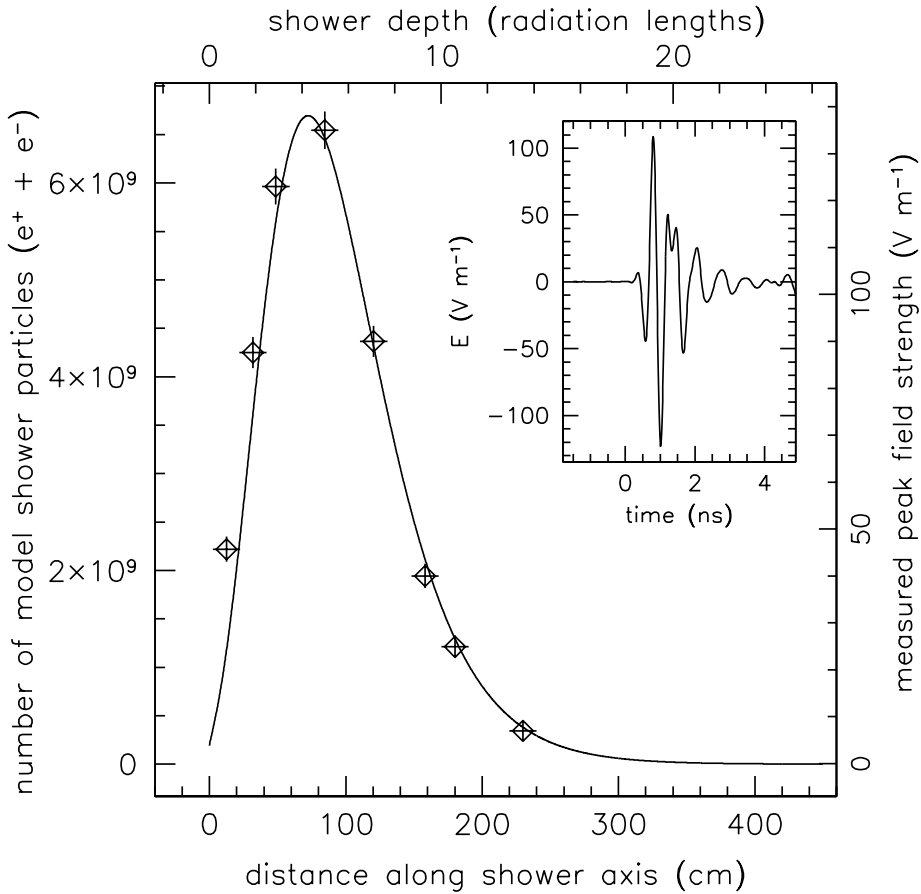}
\caption{Expected shower profile, with measured
peak field strengths plotted as diamonds, and scaled to
match the curve at about the midpoint of the measured range. 
Inset: Typical pulse time profile, 
here from the measurement near shower maximum, at about 45 cm from
the upstream edge of the target.}
\label{profile}
\end{center}
\end{figure}

Since CR propagates as a conical bow shock from the shower core, the
pulse wavefront traverses any line parallel to the shower axis at the
speed of the shower ($\simeq c$) rather than the local group velocity
$c/n$. Using our buried dipole array we have confirmed such behavior
in the pulses we observed: the arrival times of the pulse at each
dipole imply a radiative bow shock at $v/c = 1.0 \pm 0.1$, 
and are inconsistent with the measured group velocity 
($0.6~c$) in sand. 

Additional observations with an external S-band horn at the top
interface of the sand show evidence for total internal
reflection, via the $\sim 20$ dB attenuation of the pulse
transmitted through the surface, which is approximately 
parallel to the shower axis.
This latter behavior is characteristic of CR, since the Cherenkov
angle is the complement of the angle of total internal reflection,
and the radiation cannot propagate through a plane
surface parallel to the charged particle trajectory.

\begin{figure}
\begin{center}
\leavevmode
\epsfxsize=5in
\epsfbox{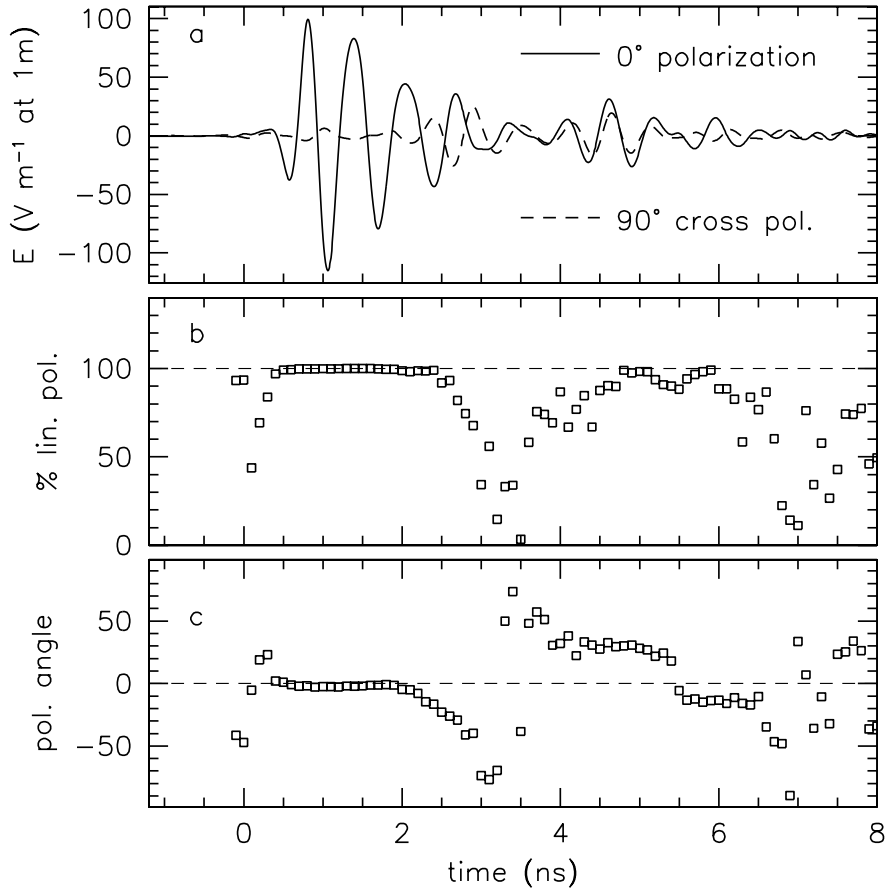}
\caption{Polarization analysis of the pulses recorded
by the S-band horns. (a) the measured field strength of the
pulse at a position corresponding to $\sim 0.5$ m beyond the
expected shower maximum. 
(b) the fractional linear polarization.
(c) the angle of the plane of polarization, where
$0^{\circ}$ is measured from the plane containing the shower axis
and the horn phase center. The leading portion of the pulse
shows polarization characteristics consistent
with Cherenkov radiation. The loss of polarization
after the first $\sim 2$~ns is due to late-arriving reflections.}
\label{polplot}
\end{center}
\end{figure}

\subsection{Polarization measurements}

Pulse polarization was measured with an S-band horn
at a single shower position, corresponding to a shower position of
0.5 m past the expected shower maximum.  Field intensities
were measured with the horn rotated at angles 0, 45, 90 and 135 degrees
from horizontal. 
Fig. 3(a) shows the pulse profile for both the $0^{\circ}$ 
and $90^{\circ}$ (cross-polarized) orientations of the
horn. Fig. 3(b) and 3(c) show the derived degree of linear
polarization and the angle of the plane of polarization,
respectively. In both respects the results are completely
consistent with Cherenkov radiation from a source along the shower
axis. Since the position was somewhat downstream of shower
maximum, reflections from the upper surface of the sand
also enter the pulse profile at later times; these are
clearly evident in the time series and as a resulting loss of polarization
beyond $\sim 2$~ns.

\subsection{Coherence \& radio frequency dependence of field strength}

In Fig. 4 we plot the measured dependence of the field strengths
as a function of the total energy of the showers and the 
radio frequency band for different antennas. Fig. 4(a)
shows a typical sequence of pulse field strengths, in this case
for the external S-band horn, versus the total shower energy,
which was varied both by changing the beam current and the thickness
of the bremsstrahlung radiators.
The dot-dash curve is a least-squares power-law fit to the
data, of the form 
$|{\bf E}| ~\propto~ (W_{T})^{\alpha}$
where  $|{\bf E}|$ is the electric field strength and 
$W_{T}$ is the total
shower energy. The fit yields $\alpha= 0.96\pm 0.05$, consistent
with complete coherence of the
radiation, implying the characteristic
quadratic rise in the corresponding pulse power with shower energy.

Fig. 4(b) shows the spectral dependence of the radiation,
which is consistent with the linear rise with frequency that
is also characteristic of Cherenkov radiation.
Fig. 4(b) also plots a curve based on a semi-empirical parameterization
\cite{ZHS92,Alv00}. This parameterization is based in part on 
analytic methods which explicitly treat the distribution of the 
shower charge excess, and have been verified to yield the proper 
Frank-Tamm\cite{Tam39}
results in limiting cases. This analysis is combined with simulations of
radio emission for TeV showers in ice to determine the
empirical parameters. Here we adapt these results from ice to silica sand, 
which has similar RF properties. The resulting field strength can
be written in the 
form~\cite{ZHS92,Alv00}: 
\begin{equation}
R~|{\bf E}| ~=~ A_0 K \epsilon 
\left ( {W_{T} \over 1~{\rm TeV}} \right )
{\nu \over \nu_0} \left ( {1 \over 1 + 0.4(\nu/\nu_0)^{\delta} } \right ) ~
({\rm V ~m^{-1}~ MHz^{-1}} ),
\end{equation}
where  $R$ is the distance to the source, 
$\nu$ is the radio frequency, and 
the decoherence frequency $\nu_0 = 2500$ MHz for silica sand
($\nu_0$ scales mainly by radiation length).
The leading coefficient is $A_0 = 2.53 \times 10^{-7}$,
and $\delta=1.44$ (both are determined empirically)\cite{Alv00}. 
Here $W_T = N_e \eta W_e$ gives the shower energy as the
product of the total number of electrons $N_e$, the thickness $\eta$ of the
radiator in radiation lengths, and the electron energy $W_e$.
The factor $K = 0.47$ takes into account the 
difference in radiation lengths and density between ice
and silica sand,
which reduces the overall tracklength.  Finally,
$\epsilon \simeq 0.5$ accounts for the fact that the
limited antenna apertures are only sensitive to $\sim 1/2$ of the total 
field strength. There is also additional uncertainty due to near-field 
effects, which we discuss further below. The uncertainties are 
estimates of the combined
systematic and statistical uncertainty.  Note that
Fig. 4(b) compares absolute field strength measurements to the predictions
and the agreement is very good.
\begin{figure}
\begin{center}
\leavevmode
\epsfxsize=5in
\epsfbox{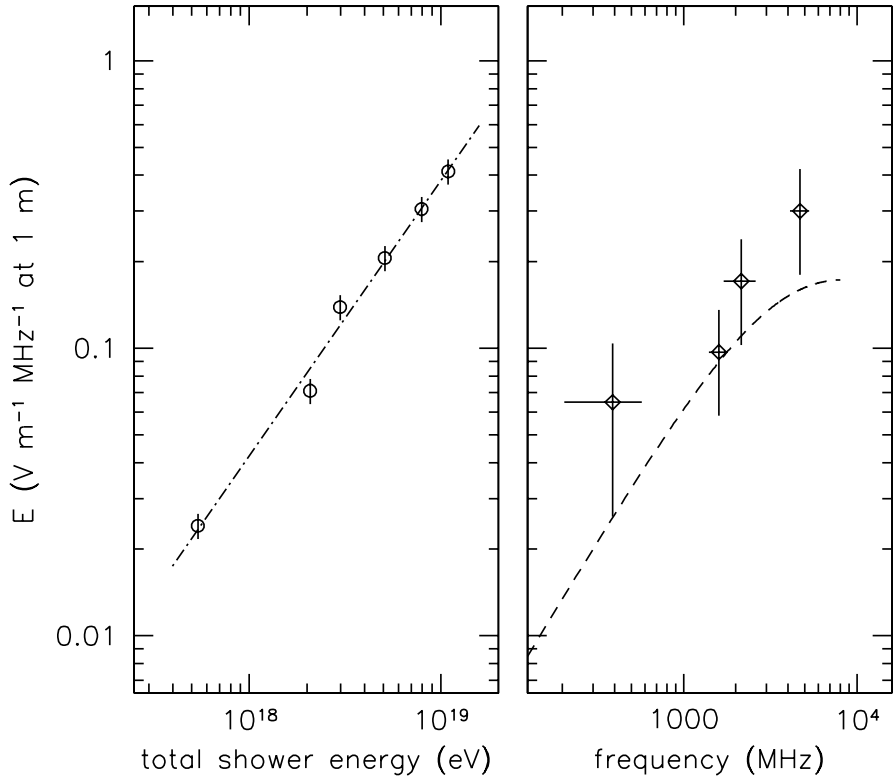}
\caption{ (a) The measured coherence
of the pulse electric field strength with change in 
shower energy, at 2.1~GHz. The curve indicates a least squares fit
to the data, and only the statistical errors
are shown. (b) The spectral dependence of the measured
pulse field strengths. The horizontal bars indicate the
bandwidth of the antenna used, and the vertical bars the
combined systematic and statistical uncertainties. 
The dashed curve is
the prediction based on a Monte Carlo simulations of showers
in ice scaled to silica sand.}
\label{cohpower}
\end{center}
\end{figure}

\section{Discussion}

We have demonstrated that the radiation we have observed is
coherent, and 100\% linearly polarized. The plane of polarization
coincides with the plane containing the antenna and shower axis.
The radiation is pulsed with time durations much shorter than
the inverse bandwidth of our antennas. The strength of the
pulse is strongly correlated to the size (in particle number) of
the shower region that appears to produce it. 
All of these observed characteristics are consistent with the
hypothesis that what we have observed is coherent Cherenkov radiation.
Because of the strong correlation with the shower profile, and the
physical constraint that a shower with no net excess charge
cannot radiate, we conclude that excess charge production
along the shower is the source of the propagating charge in
the silica dielectric which leads to the CR observed.
These conclusions are strengthened by the fact that the 
pulse strength is consistent with Monte Carlo predictions for
such showers.

Our observations are inconsistent with radiation from geomagnetic
charge separation as observed in extensive air showers. The
most striking evidence for this is the fact that the plane of
polarization is clearly aligned with the shower axis rather than the
local geomagnetic dip angle ($62^{\circ}$). Given the
approximate east-west orientation of the shower axis, boosted
dipole radiation from geomagnetic charge separation should
produce an electric field polarization with significant components
orthogonal to what we have observed.

We note that in all cases our measurements are likely to
be made in near-field conditions, and we have not attempted
to correct for these effects. Recent
studies\cite{Bun00,Alv00} have begun to treat these issues, but
are not straightforward to apply in our case. In any case, near
field effects should generally {\em decrease} the measured field
strengths relative to far field measurements. 

The total energy of our cascades is as high as $10^{19}$ eV,
but these showers consist of the superposition of many
lower-energy showers. Higher-energy effects\cite{Alv98,Alv99}
that elongate the shower
during its development are not present. These effects all tend
to increase the total tracklength of the shower, at the expense
of a lower instantaneous particle number density.
The net effect is that the total radiated power is 
expected to be approximately
conserved, but the angular spectrum can change significantly,
with a predicted sharpening of the Cherenkov angular distribution
for high energy showers. We have not measured the angular
spectrum of the radio pulses, and thus any scaling of our results
to high energy showers should consider corrections for these
effects, which are likely to increase the peak field strength in
far-field measurements. 

Since the field strength scales linearly in shower energy
and inversely with distance from the source, extrapolations to
determine the energy threshold of existing experiments is straightforward,
after correcting for the differences in material properties.
For Antarctic ice experiments, the use of the existing simulations
\cite{ZHS92,Alv98,fri96,Alv00} appears completely justified. For experiments
observing the lunar regolith\cite{Zhe88,Dag89,Han96,Gor99}, 
silica sand shares many similarities
with the lunar surface material, and the expected cascade energy threshold,
by direct scaling of our results, 
is $\sim 2 \times 10^{19}$ eV,
and may be somewhat lower depending on the angular intensity
effects discussed above. 

We conclude that, in combination with our previous measurements
of coherent RF transition radiation~\cite{P1}, these results
have established a firm experimental basis for radio-frequency
detection of high energy cascades in solid media, 
either through interaction
within a dielectric (for CR), or via passage through dielectric 
interfaces (for TR). Above cascade energies of $\sim 10^{16}$~eV, 
these secondary emission processes become dominant over
others (for example, optical Cherenkov or
fluorescence emission) in the number of 
quanta produced\cite{Mark86}.
Thus experiments designed to exploit this effect
in the detection of ultra-high energy particles
can now be pursued with even greater confidence.

\section*{Acknowledgments}
We thank D. Besson, R. Rose, L. Skjerve, and M. Spencer for
the loan or construction of various antennas used in the experiment.
We thank the members of the SLAC accelerator and EF departments
for invaluable assistance before, during and after the run.
This research was supported in part by 
by DOE contract DE-FG03-91ER40662 at UCLA and by the A. P. Sloan Foundation.
It has been performed in part at the Jet Propulsion
Laboratory, California Institute of Technology, under contract with
the National Aeronautics and Space Administration.
The Stanford Linear Accelerator Center is supported by
the U.S. Department of Energy, with work performed under
contract DE-AC03-76SF00515.
Argonne support came from the U.~S. Department
of Energy, Division of High Energy Physics contract 
W-31-109-ENG-38.

\end{document}